\documentclass[final,3p,times]{elsarticle}

\usepackage{natbib}
\usepackage{array}
\usepackage{booktabs}
\usepackage{tabu}
\usepackage{dcolumn}
\usepackage{amsmath}
\usepackage{amsfonts}
\usepackage{amssymb}
\usepackage{graphicx}
\usepackage{subfigure}
\usepackage{epsfig}
\usepackage{epstopdf}
\usepackage{xcolor}
\usepackage{float}
\usepackage[breaklinks=true,colorlinks=true,linkcolor=blue,urlcolor=blue,citecolor=red]{hyperref}

\usepackage{dcolumn}
\usepackage{bm}

\def\be{\begin{equation}}
\def\ee{\end{equation}}
\def\bea{\begin{eqnarray}}
\def\eea{\end{eqnarray}}

\journal{Physics of The Dark Universe}

 \rm
\makeatletter
\newcommand*{\shifttext}[2]{%
	\settowidth{\@tempdima}{#2}%
	\makebox[\@tempdima]{\hspace*{#1}#2}%
}
\makeatother
 
\makeatother

\newcommand{\mergv}{%
	\mathop{%
		\rlap{$\mathbb{V}$}%
		\mkern2mu
		\shifttext{2.5pt}{$\mathbb{V}$}%
	}%
}

\begin{document}

\begin{frontmatter}

\title{ \bf \Large {Observational Constraints on the Dark Energy with a Quadratic Equation of State}}

\author[IPM]{Hossein Moshafi\corref{cor1}}
\ead{moshafi86@gmail.com;moshafi@ipm.ir}

\author[IPM]{Alireza Talebian}
\ead{talebian@ipm.ir}

\author[IPM,IMRC]{Ebrahim Yusofi}
\ead{eyusofi@ipm.ir}

\author[SMS]{Eleonora Di Valentino}
\ead{e.divalentino@sheffield.ac.uk}

\cortext[cor1]{Corresponding author}

\address[IPM]{School of Astronomy, Institute for Research in Fundamental Sciences (IPM),  P. O. Box 19395-5531, Tehran, Iran}

\address[IMRC]{Innovation and Management Research Center, Ayatollah Amoli Branch, Islamic Azad University, Amol, Mazandaran, Iran}

\address[SMS]{School of Mathematics and Statistics, University of Sheffield, Hounsfield Road, Sheffield S3 7RH, United Kingdom}


\begin{abstract}
In this study, we introduce a novel late-time effective dark energy model characterized by a quadratic equation of state (EoS) and rigorously examine its observational constraints. Initially, we delve into the background dynamics of this model, tracing the evolution of fluctuations in linear order. Our approach involves substituting the conventional cosmological constant with a dynamically effective dark energy fluid. Leveraging a diverse array of observational datasets encompassing the Planck 2018 Cosmic Microwave Background (CMB), Type Ia Supernovae (SNe), Baryon Acoustic Oscillations (BAO), and a prior on the Hubble constant $H_0$ (R21), we constrain the model parameters. We establish the model's consistency by comparing the Hubble parameter as a function of redshift against observational Hubble data (OHD), benchmarking its performance against the Standard $\Lambda$CDM model. Additionally, our investigation delves into studies of the model's dynamical behavior by computing cosmological parameters such as the deceleration parameter, relative Hubble parameter, and the evolution of the Hubble rate. Furthermore, employing Bayesian analysis, we determine the Bayesian Evidence for our proposed model compared to the reference $\Lambda$CDM model. While our analysis unveils the favorable behavior of the model in various observational tests, the well-known cosmological tensions persist when the full dataset combination is explored.

\end{abstract}


\end{frontmatter}
\tableofcontents

\

\section{Introduction}
One of the most groundbreaking discoveries in recent cosmology is the observation of cosmological acceleration in our Universe. Despite this revelation, the true nature of this acceleration remains elusive and is referred to as dark energy (DE). According to the widely accepted standard model of cosmology, called $\Lambda \rm{CDM}$ (which aligns remarkably well with current cosmological observations), DE is a cosmological constant and accounts for $\sim 70\%$ of the energy budget of the Universe. For a more comprehensive understanding of DE, various cosmological models have been introduced.
Constructing a cosmological model requires two essential elements: a gravitational theory defining the Universe's geometry and a comprehensive picture of its matter content. In this context, DE can be conceptualized either as an altered form of matter (such as quintessence, k-essence, phantom, $\cdots$) or as a modification within gravitational theories (like $f(R)$ gravity, Gauss–Bonnet, $\cdots$)~\citep{amendola2010dark}.

The models based on modified matter can be achieved through the introduction of an exotic matter source exhibiting negative pressure. The most straightforward instances include fluids characterized by a linear equation of state (EoS)~\citep{Harko:2022unn,Li:2019yem,Moshafi:2022mva,Mostaghel:2016lcd,Mostaghel:2018pia,Yang:2020zuk}, denoted as $p= w \rho$, where $p$ represents pressure and $\rho$ signifies energy density.


In diverse cosmological models, including those rooted in extra dimensions (such as String theory, Brane-worlds, etc.), and within intricate and more realistic systems, the relationship between pressure and density may not adhere to a linear correlation~\citep{Wood:2023lis,Jalalzadeh:2023upb,Cicoli:2023opf,Danielsson:2022lsl}. If we assume $p=p(\rho)$ for any barotropic fluid to be an analytic function, we can consider a more general form of the equation of state of the form $p=p_0 + A_1 \rho + A_2 \rho^2 + \mathcal{O}( \rho^3)$. In the Brane-world~\citep{Shiromizu:1999wj,Bridgman:2001mc,Langlois:2003yy} scenario, the assumption of the extra dimensions produces a quadratic term in the energy density. In addition, it has been shown that Loop Quantum Gravity (LQG) corrections result in a modified Friedmann equation with the modification appearing as a negative term which is quadratic in the energy density~\citep{Vandersloot:2005kh,Ashtekar:2006uz,Ashtekar:2006wn,Saaidi:2012qp,Sengupta:2023sua}.

In this paper, we study an effective dark energy fluid with a quadratic EoS~\citep{Ananda:2005xp}. The non-linear term in EoS can be originated from the merger process of clusters/voids and may act as a possible source for cosmic acceleration~\citep{Yusofi:2022hgg, Mohammadi:2023idz}. The model can explain the evolution of voids~\citep {Sheth:2004mnr, WEYGAERT:2011csd} in an under-dense environment within a void-dominated universe~\citep{Redmount:1988rih}. Also, it is shown that this fluid has a possible balancing effect on the cosmological expansion~\citep{Mohammadi:2023idz}. In this work, we substitute a new energy density component $\rho_{\rm v}$, instead of the cosmological constant $\rho_{\rm \Lambda}$ 
in $\Lambda$CDM, and we call it $\mergv$CDM model. We consider the observational data (including the deceleration parameter, relative Hubble parameter, and the evolution of the Hubble rate) in the framework of the proposed model. 

For comprehensive evaluation against other cosmological models, examining how the $\mergv$CDM model performs with respect to the cosmological tensions, including the Hubble constant $H_0$ tension~\citep{Verde:2019ivm, Riess:2019qba,DiValentino:2020zio}, proves to be particularly insightful. There was another well-studied problem called $S_8$ tension~\citep{KiDS:2020suj,DES:2021bvc,DES:2022ccp,Planck:2018vyg,Li:2023tui,Dalal:2023olq,DiValentino:2020vvd} which seems to be alleviated by the recent study by~\cite{Kilo-DegreeSurvey:2023gfr}. 
Several models have been proposed for alleviating these problems in the literature~\citep{Knox:2019rjx,Jedamzik:2020zmd,DiValentino:2021izs,Abdalla:2022yfr,Kamionkowski:2022pkx,Escudero:2022rbq,Vagnozzi:2023nrq,Verde:2023lmm,Khalife:2023qbu,Moshafi:2022mva,Moshafi:2020rkq,Khosravi:2017hfi,Banihashemi:2020wtb, Talebian:2023lkk}. Roughly speaking, two primary categories of DE models aimed at addressing the $H_0$ tension by introducing novel physics: early-time and late-time DE models. These categories depend on the modifications to the expansion history, whether they occur before or after the recombination epoch, respectively. Within this framework, the proposed $\mergv$CDM model in our study aligns with the latter category.

The paper is organized as follows: in Section~\ref{sec:model} we explain the dynamics of the model and 
derive the equation of motion of linear perturbations. For testing the model, we introduce the observational data sets and the statistical methods used in this work in Sections~\ref{sec:obs} and \ref{sec:stat}, respectively. Our results for this model in the light of the combination of different datasets are summarized in Section~\ref{sec:results}. Finally, we discuss the results in Section~\ref{sec:Discussion}.

\section{The Model with a Quadratic EoS}
\label{sec:model}

We consider a model consisting of a new variable energy density $\rho_{\rm v}$ instead of the constant dark energy density
in the standard $\Lambda{\rm CDM}$ model. The new source is a cosmological barotropic fluid with a non-linear EoS. In this work, we are focusing on the quadratic case~\citep{Ananda:2005xp, Khanpour:2017das, Mohammadi:2023idz}, 
\begin{align}
\label{P_v}
p_{\rm v} = w_{\rm v}~ \rho_{\rm v} + \beta~\rho_{\rm v}^2
\end{align} 
in which $P_{\rm v}$ and $w_{\rm v}$  are the pressure and the equation of state parameter, respectively. The new parameter $\beta$ comes from the non-linearity effect in the over-dense regime that can originate from the merger of clusters/voids in this area and sets the characteristic energy scale $\rho_{\rm c}$
of the quadratic term
\begin{align}
\beta \equiv \dfrac{b}{\rho_{\rm c}}
\end{align}
where $b$ is a constant non-zero parameter that must be determined with the observational data. 

As an example of application, the equation of state of~\eqref{P_v} can describe a cosmic fluid with two different regimes: under-dense regime (linear term $\propto \rho_{\rm v}$) and over-dense regime (non-linear term $\propto \rho_{\rm v}^2$). Our model requires the existence of some phenomenological interactions as the possibility of merging and collapsing of these two regimes. More precisely, starting with the linear EoS for the fluid, the merging (collapsing) of two under-dense (over-dense) regimes leads to the final pressure of the fluid being modified by the non-linear term in \eqref{P_v}~\citep{Khanpour:2017das}. In the general case, the rate of merging and collapsing could be a function of space and time, and two under-dense and over-dense regions can coexist at the same time~\citep{Shim:2020wyj}.

This picture is consistent with the findings on the evolution and merging of the vast cosmic voids~\citep{Sheth:2004mnr, WEYGAERT:2011csd, Redmount:1988rih} if voids are considered as the under-dense regime of our fluid while the density of the fluid on the boundary of voids where the dark matter halos are located is high and non-linear.
More specifically, in the evolution of voids, two processes are reported: {\textit{merging}} of two voids into larger voids in large-scale under-dense regions and their \textit{collapsing} and disappearance in local over-dense regions~\citep{Sheth:2004mnr}. For instance, two small voids merge and form a giant cosmic void. When the adjacent sub-voids meet up and merge, the matter between them is squeezed into thin walls and filaments~\citep{WEYGAERT:2011csd}. In such a merging cosmic fluid, our universe consists of under-dense and over-dense regions in the form of a cosmic web, in which under-density regions are located inside the voids, and over-dense regions including strings, clusters, and shells are located at the boundary of the voids~\citep{Redmount:1988rih}.

In our model, we assume the characteristic energy scale $\rho_{\rm c}$ be the critical energy density of the Universe, i.e. $\rho_{\rm c} \sim \rho_0$. This assumption is based on the previous studies~\citep{Yusofi:2022vsg, Yusofi:2022hgg} in which the non-linear term in the total pressure can be related to the surface tension of the voids. 


Using the \textit{continuity equation}, the energy density of the fluids $\rho_{\rm v}$ as a function of the scale factor $a$ for the cases $w_{\rm v} \neq -1$  is given by~\citep{Khanpour:2017das,Mohammadi:2023idz}
\begin{equation}
\label{eq:rov}
\rho_{\rm v}(a)= \frac{\rho_{\rm v_0}a^{-3(1+w_{\rm v})}}{1+\frac{\beta\rho_{\rm v_0}}{1+w_{\rm v}}[1-a^{-3{(1+w_{\rm v})}}]}\,,
\end{equation}
and for the special case ($w_{\rm v} = -1$) is given by
\begin{equation}
\label{eq:rov2}
\rho_{\rm v}(a)= \frac{\rho_{\rm v_0}}{1+3\beta\rho_{\rm v_0}\ln{a}},  
\end{equation}
where $\rho_{\rm v_0}$ is the energy density of the fluid at present $a=1$. In the limit $\beta\rightarrow0$, the model \eqref{eq:rov} covers the phantom ($w_{\rm v}<-1$)~\citep{Caldwell:1999ew} and quintessence ($w_{\rm v} \geqslant -1$) models~\citep{papantonopoulos2005physics}  while the model \eqref{eq:rov2} meets the well-known $\Lambda{\rm CDM}$ model. 

As seen, the model is a dynamical dark energy in which the cosmological constant is replaced by a quadratic barotropic fluid. The cosmic background evolution is described through the Friedmann equation,
\begin{equation}
\label{eq:Hubble-parameter}
H^2(z) = \frac{8  \pi G}{3} \sum_i \rho_i(z)
\end{equation}
in which $G$ is the gravitational constant and $H$ is Hubble parameter and the total energy density consists of
\begin{equation}
\sum_i \rho_i(z) =  \rho_{\rm r} (z) + \rho_{\rm b} (z) + \rho_{\rm cdm} (z) + \rho_{\rm v}(z) 
\label{eq:densities-sum}
\end{equation}
where $\rho_{\rm r}$, $\rho_{\rm b}$ and $\rho_{\rm cdm}$ are the energy density of radiation, baryon, and cold dark matter, respectively, while $\rho_{\rm v}$ is given by \eqref{eq:rov} or \eqref{eq:rov2} depending on $w_{\rm v}$. In the following, we study some important cosmological parameters in the framework of the $\mergv$CDM model, by careful examination of the background evolution and first-order perturbations in the light of observational data.

To investigate the linear growth of density perturbations of both the pressureless dark matter and dark energy fluid in our model, we use perturbed Einstein's equation in the conformal-Newtonian gauge. In this formalism, the linear density contrast of non-relativistic DM, $\delta_{\rm m}$, and DE, $\delta_{\rm v}$, satisfy the following equations in Fourier space~\citep{Amendola:2015ksp}
\begin{align}
 k^2 \Phi+ 3 \mathcal{H} \left( \Phi^\prime - \mathcal{H} \Psi \right) &= 4 \pi G a^2 \left[ \rho_{\rm m} \delta_{\rm m} + \rho_{\rm v} \delta_{\rm v} + \rho_{\rm r} \delta_{\rm r} \right],
\\
 k^2 \left(\Phi^\prime - \mathcal{H} \Psi \right) &= -4 \pi G a^2 \left[ \left(1+w_{\rm m} \right) \rho_{\rm m} \theta_{\rm m} + \left(1+w_{\rm v}+ \beta \rho_{\rm v} \right) \rho_{\rm v} \theta_{\rm v} + \left(1+w_{\rm r}  \right) \rho_{\rm r} \theta_{\rm r} \right],
 \\
 \Psi &= - \Phi,
 \\
 \Phi^{\prime \prime} + 2 \mathcal{H} \Phi^\prime - \mathcal{H} \Psi^\prime - \left( \mathcal{H}^2 + 2 \mathcal{H}^\prime \right) \Psi &= -4 \pi G a^2 \left[ c^2_{\rm s,m}  \rho_{\rm m} \delta_{\rm m}+ \left( w_{\rm v} + 2\beta \rho_{\rm v} \right)  \rho_{\rm v} \delta_{\rm v} + c^2_{\rm s,r} \rho_{\rm r} \delta_{\rm r} \right],
 \\
 \delta^\prime_{\rm v} + 3 \mathcal{H} \beta \rho_{\rm v} \delta_{\rm v} &= - \left( 1+ w_{\rm v} + \beta \rho_{\rm v} \right) \left( \theta_{\rm v} + 3 \Phi^\prime \right),
 \\
 \theta^\prime_{\rm v} + \theta_{\rm v} \left[ \left(1 - 3 w_{\rm v} \right) \mathcal{H} - 6 \beta \rho_{\rm v} \right] &= k^2 \left( \frac{w_{\rm v} + 2\beta \rho_{\rm v}}{1+w_{\rm v} + \beta \rho_{\rm v}} \delta_{\rm v} + \Psi \right)
\end{align}
where $\eta$ is the conformal time and the prime denotes the derivative to the conformal time.  $\Psi$ \& $\Phi$ are Bardeen potentials and $\mathcal{H}$ is conformal Hubble parameter. $c^2_{\rm s,r}$ and $c^2_{\rm s,m}$ are the sound speeds and $\theta_{\rm v}, \theta_{\rm m}$ and $\theta_{\rm r}$ are divergence of velocities of corresponding components. Finally $\rho_{\rm m}=\rho_{\rm b}+\rho_{\rm cdm}$ and $\rho_{\rm r}=\rho_\gamma+\rho_\nu$ are the density matter and density radiation, respectively. 
The power spectrum of matter perturbations $P(k)$ is defined as
\begin{equation}
\label{eq:power-spectrum-defination}
\langle {\delta}_{\rm m}(\vec{k})~ {\delta}_{\rm m}(\vec{k^\prime}) \rangle = \left(2 \pi \right)^3 \delta_D \left( \vec{k} + \vec{k^\prime} \right) P(k)
\end{equation}
where $\delta_D$ is a Dirac delta function and ${\delta}_{\rm m}$ is Fourier transformed density field. The variance of the mass distribution smoothed on $R=8 h^{-1} \rm{Mpc}$ scales is defined
\begin{equation}
\label{eq:sigma8-definition}
\sigma_8^2 = \frac{1}{2 \pi^2} \int P(k) \left[ \frac{3j_1(kR)}{kR} \right]^2 k^2 dk
\end{equation}
where $ j_1(kR)$ is a spherical Bessel function. 

We've implemented the background and the perturbation equations with correct modifications into the \texttt{CAMB} code~\citep{Lewis:1999bs} and calculated matter, and temperature and polarization power spectra.

\subsection{Redshift Dependence of the Equation of State and the Speed of Sound}

\begin{figure}[ht]
\centering
  \includegraphics[width= .7\linewidth]{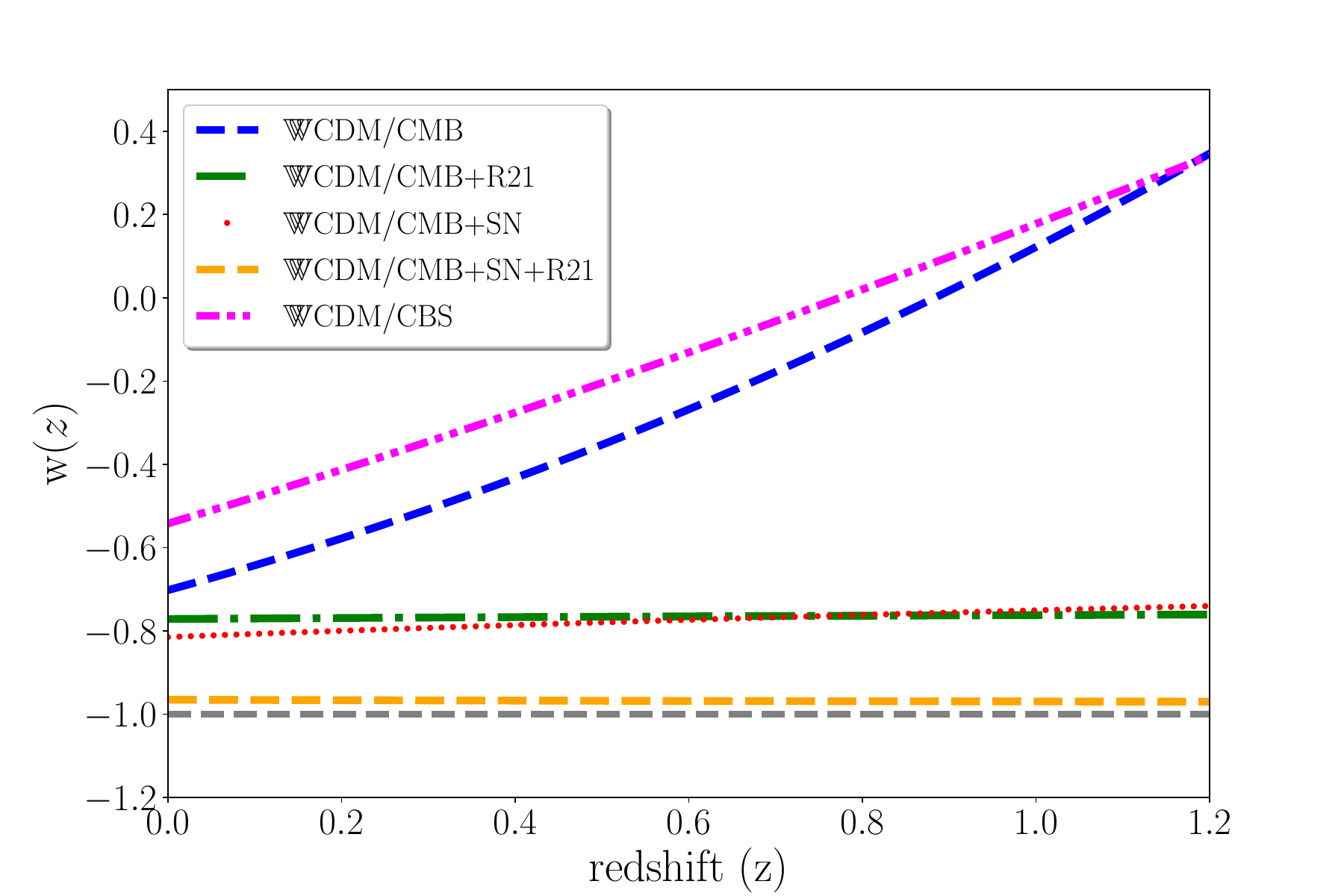}
    \includegraphics[width= .7\linewidth]{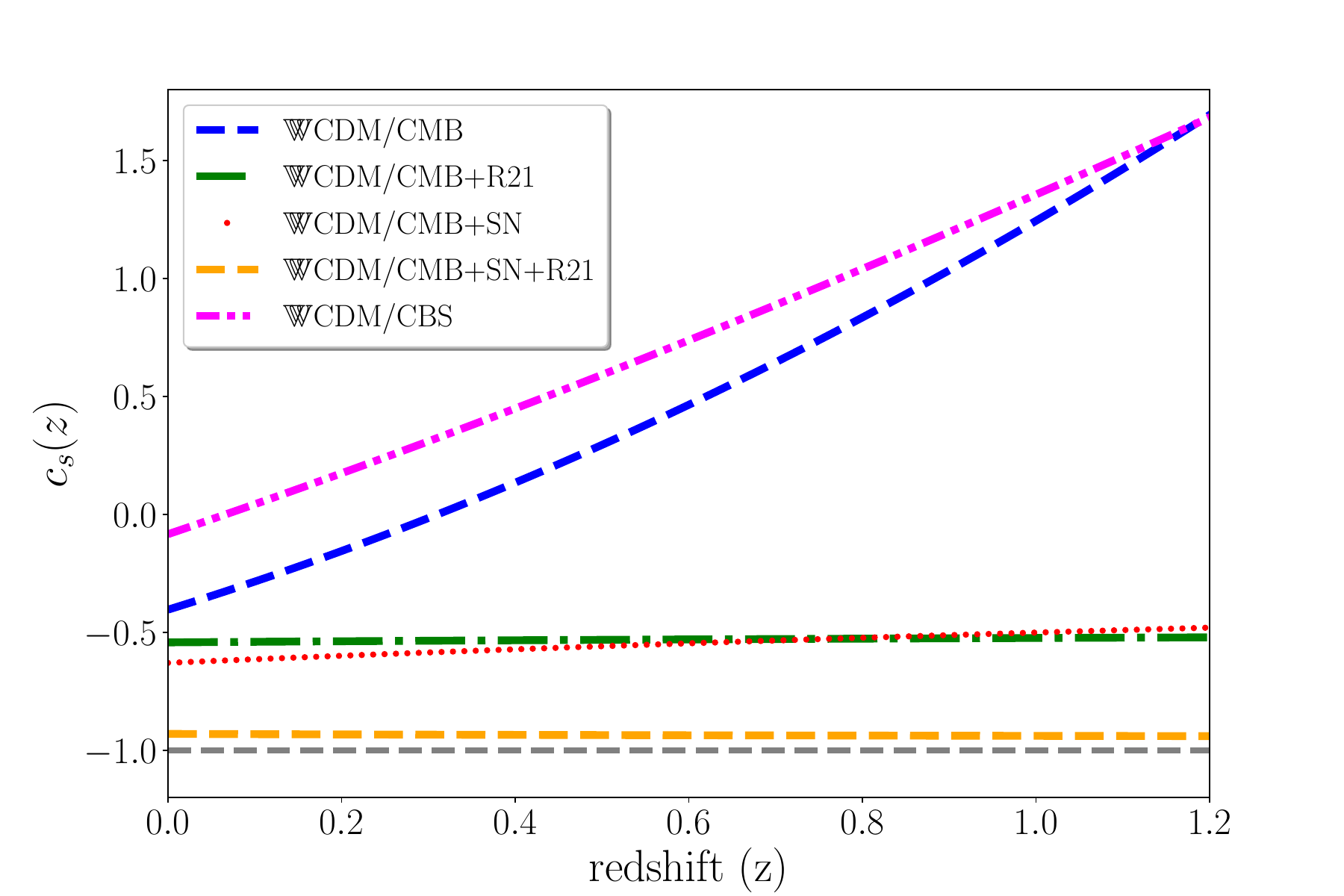}
  \caption{\footnotesize  Equation of State and Sound Speed of Dark Energy for different combination of data.}
  \label{fig:w-cs}
\end{figure}

The equation of state parameter $w$ quantifies the pressure-to-energy density ratio and the sound speed $c_s^2$ measures how quickly small perturbations can propagate through a fluid, crucial for understanding structure growth in the universe.
The generalized forms for $w$ and $c_s^2$ can be write as,
\begin{align}
	\label{cvt4}
	w(\rho) &= \frac{p}{\rho},
	\\
	\label{cvt44}
	c_s^2(\rho) &= \frac{dp}{d\rho} \,.
\end{align}
For quadratic case \eqref{P_v}, we obtain
\begin{align}
	\label{cvt5}
	w(\rho_{\rm v})&=w_{\rm v}+\beta\rho_{\rm v} \,	
	\\
	\label{cvt55}
	c_s^2(\rho_{\rm v})&=w_{\rm v}+2\beta\rho_{\rm v} \,.
\end{align}
Since the energy density depends on redshift (as seen from \eqref{eq:rov} and \eqref{eq:rov2}), the equation of state and sound of speed also being dependent on the redshift $z$ as follows
\begin{align}
	\label{cvt6}
	w(z) &=w_{\rm v}+\beta\rho_{\rm v}(z)~,
	\\
	c_s^2(z) &=w_{\rm v}+2\beta\rho_{\rm v}(z)
\end{align}
In Fig.\ref{fig:w-cs}, we have illustrated the evolution of equation of state and the speed of sound in terms of the redshift.

\section{Observational data}
\label{sec:obs}

In this section, we describe the cosmological data sets used in this work in brief. The description of the data catalogs is as follows:
\begin{enumerate}

\item {\bf CMB}: We use the most precise full-sky measurements of Cosmic Microwave Background (CMB) radiation performed by \emph{Planck} satellite. We use both high-$\ell$ temperature and polarization angular power spectra from the final release of  ``\emph{Planck 2018}'' baseline \texttt{PLIK-LITE TTTEEE} along with \emph{Planck} \texttt{low}-$\ell$ and \texttt{low}-E \texttt{(SimAll)} ($\ell \leq 30$)~\citep{Planck:2018vyg, Planck:2018lbu, Planck:2019nip}. We mention all of \emph{Planck} data (including temperature and polarization) by ``CMB''.

\item{\bf R21}: To test the ability of this model to reconcile the $H_0$ tension, we additionally include a Gaussian prior of the form $H_0=73.04 \pm 1.04$ $\rm{km}/ \rm{s}~ \rm{Mpc}$, as reported by the SH0ES Collaboration~\citep{Riess:2021jrx}. We refer to this prior as ``R21''.

\item {\bf BAO}: We also consider the various measurements of the Baryon Acoustic Oscillations (BAO) from different galaxy surveys as the \emph{Planck} collaboration in their 2018 analysis~\citep{Planck:2018vyg}, i.e. 6dFGS~\citep{Beutler:2011hx}, SDSS-MGS~\citep{Ross:2014qpa}, and BOSS DR12~\citep{BOSS:2016wmc}. We mention all these data points by ``BAO''.

\item {\bf SN}: We include the measurements of  the $1048$ Supernovae Type Ia luminosity distance in the red-shift interval $z \in [0.01, 2.3]$, from the Pantheon sample~\citep{Pan-STARRS1:2017jku}. We show this catalog of SuperNovae by ``SN''.

\end{enumerate}
In some analysis, we use a combination of ``CMB+BAO+SN'' data which we refer to this combination as ``CBS''.

\section{Statistical methods and Analysis}
\label{sec:stat}

Here, we briefly introduce the statistical methods used in our analysis. The cosmological analysis we carry out in this work is based on Bayesian inference. To analyze the data and extract the constraints on the cosmological parameters for $\mergv$CDM model, we have modified carefully the widely used cosmological Markov Chain Monte Carlo package \texttt{CosmoMC}~\citep{Lewis:2002ah, Lewis:2013hha}, which is publicly available.\footnote{http://cosmologist.info/cosmomc}
This package is equipped with a convergence diagnostic based on the Gelman and Rubin statistic~\citep{Gelman:1992zz}, assuming $|R-1| < 0.01$ for all parameters, and implements an efficient sampling of the posterior distribution using the fast/slow parameter decorrelations~~\citep{Lewis:2013hha}.

One measure to check the consistency of our model with observational data in comparison to the standard model is the ``Akaike Information Criterium'' (AIC) which is a tool to measure the improvement of the fit. We compute the AIC of the extended model $\mergv$CDM relative to that of $\Lambda$CDM, defined as
 \begin{equation}
    \Delta {\rm AIC} = 
    \chi^2_{\mathrm{min},{\mergv \rm{CDM}}} - \chi^2_{\mathrm{min},\Lambda \rm CDM} + 2 (N_{\mergv \rm{CDM}}-N_{\Lambda \rm CDM})~, 
    \end{equation} 
   where $N_{\mergv \rm{CDM}}$ and $N_{\Lambda \rm CDM}$ stand for the number of free parameters of $\mergv$CDM model and $\Lambda$CDM model~\citep{1100705}, respectively.
   
To put the model to the test for Hubble tension we use the ``rule of thumb difference in mean'' or Gaussian Tension (GT)~\citep{Raveri:2018wln}, defined as

    \begin{equation}
    \frac{{\bar x}_{\mathcal{D}} - {\bar x}_{\rm SH0ES}}{({\sigma}_{\mathcal{D}}^2 + {\sigma}_{\rm SH0ES}^2)^{1/2}} \, ,
    \label{eq:GT}
    \end{equation}
where $\overline{x}_{\mathcal{D}}$ and $\sigma_{\mathcal{D}}$ are the mean and standard deviation of the $H_0$ for model based on observational data $\mathcal{D}$, and $\overline{x}_{\rm SH0ES}$ and $\sigma_{\rm SH0ES}$ correspond to $H_0=73.04 \pm 1.04$~$\rm{km}/ \rm{s}~ \rm{Mpc}$~\citep{Riess:2021jrx}.
\begin{table*}[t]
	\centering
	\begin{tabular} {c||c }
		Parameter & Priors      \\
		\hline
		\hline
		{$w_{\rm v}$} & $[-1.5, +1.5]$    \\
		{$b$} & $[0.0, 1.0]$   \\
		{$\Omega_b h^2$} & $[0.005, 0.1]$    \\
		{$\Omega_c h^2$} & $[0.001, 0.99]$    \\
		{$\tau_{\rm re}$} & $[0.01, 0.8]$    \\
		{$100 \Theta_{\rm MC}$} & $[0.5, 10]$    \\
		{$\log\left[10^{10} A_s \right]$} & $[1.6, 3.91]$    \\
		{$n_s$} & $[0.8, 1.2]$    \\
		\hline
	\end{tabular}
	\caption{Flat priors used on various free parameters of $\mergv$CDM model, during statistical analysis.}
	
	\label{tab:priors}
\end{table*}

We do a Maximum Likelihood Analysis by using \texttt{CosmoMC} as a cosmological MCMC code. To find the best fit of parameters of $\mergv$CDM model, we assume a parameter space as:
\begin{eqnarray}
\mathcal{P}_0 \equiv\Bigl\{w_{\rm v}, b, \Omega_{b}h^2, \Omega_{c}h^2, 100\theta_{\rm MC},
\tau, n_{s}, \ln[10^{10}A_{s}] \Bigr\}~,
\label{eq:VCDM}
\end{eqnarray}
where $b$ is the void merging parameter, $w_{\rm v}$ represents the equation of state parameter, $\tau_{\rm re}$ is the reionization optical depth, $n_s$ is the scalar spectral index, $A_{s}$ is the amplitude of the scalar primordial power spectrum, and $\Theta_{\rm MC}$ is an approximation of $\theta_*$, which represents the angular scale of the sound horizon at decoupling. We always consider a flat Universe ($\Omega_{\rm K}=0$). The priors assumed for parameters are summarized in Table\ref{tab:priors}.

\subsection{Bayesian inference}
A natural question that arises is how the model is efficient compared to the standard $\Lambda$CDM cosmology. For this purpose, we need a statistical measure for comparison between models where the base model will be fixed as $\Lambda$CDM. This statistical comparison comes through the Bayesian Evidence. Here, we use publicly available code \texttt{MCEvidence}~\citep{Heavens:2017afc,Heavens:2017hkr} to compute the Bayesian evidence of the models.

According to Bayes' theorem, the probability of a model $M$ with a set of parameters $\Theta$, in light of the observed data $D$, is given by the  \textit{Posterior} $\mathcal{P}$: 
\begin{equation}
    \mathcal{P}(\Theta \mid D,M)= \frac{\mathcal{L}(D\mid \Theta,M)\Pi(\Theta\mid M)}{\mathcal{E}(D\mid M)},
    \label{eq:bayes}
\end{equation}

\noindent where $\mathcal{L}$ is the Likelihood function, $\Pi$ represents the set of  Priors, containing the \textit{a priori} information about the parameters of the model. $\mathcal{E}$ is the so-called Evidence, to which we pay particular attention.

For a given model $M$, the Bayesian evidence $\mathcal{E}$ is the normalizing constant in the right-hand side of Eq. (\ref{eq:bayes}). It normalizes the area under the posterior $\mathcal{P}$ to unity and is given by
\begin{equation}
  \mathcal{E}(D\mid M)= \int d\Theta \mathcal{L}(D|\Theta,M)\Pi(\Theta|M)\, . 
  \label{eq:evidence}
\end{equation}

The evidence can be neglected in model fitting, but it becomes important in model comparison. When comparing two different models $M_1$ and $M_2$ using Bayes' theorem \eqref{eq:bayes}, the ratio of posterior probabilities of the two models $\mathcal{P}_1$ and $\mathcal{P}_2$ will be proportional to the ratio of their evidence, this is
\begin{equation}
    \frac{\mathcal{P}_1(\Theta_1 \mid D,M_1)}{\mathcal{P}_2(\Theta_2 \mid D,M_2)} = \frac{\Pi_1(\Theta_1|M_1)}{\Pi_2(\Theta_2|M_2)}\frac{\mathcal{E}_1(D\mid M_1)}{\mathcal{E}_2(D\mid M_2)}\, .
\end{equation}

This ratio between posteriors leads to the definition of the \textit{Bayes Factor} $B_{12}$, which in logarithmic scale is written as
\begin{equation}
    \ln \mathcal{B}_{12} \equiv \log \left[\frac{\mathcal{E}_1(D\mid M_1)}{\mathcal{E}_2(D\mid M_2)}\right] = \ln \left[\mathcal{E}_1(D\mid M_1)\right] - \ln \left[\mathcal{E}_2(D\mid M_2)\right] \, .
    \label{BF}
\end{equation}

If $\ln \mathcal{B}_{12}$ is larger (smaller) than unity, the data favors model $M_1$ ($M_2$). To assess the strength of the evidence contained in the data, we use the empirical revised Jeffreys~\citep{Jeffreys:1961} scale, see Table~\ref{tab:jeffreys}.

\begin{table}[!h]
	\centering
\begin{tabular}{cc}
\hline\hline
$|\ln \mathcal{B}_{12}|$ & $\ \ $ Evidence for model ${M}_1$ $\ \ $ \\ \hline
$|\ln \mathcal{B}_{12}| < 1$ & Inconclusive \\ 
$1 \leq |\ln \mathcal{B}_{12}| < 2.5$ & Significant \\ 
$2.5 \leq |\ln \mathcal{B}_{12}| < 5$ & Strong \\ 
$|\ln \mathcal{B}_{12} |\geq 5$ & Decisive \\ \hline\hline
\end{tabular}%
\caption{Revised Jeffreys scale quantifying the observational viability of cosmological model $M_1$ compared to some reference model $M_2$~\citep{Kass:1995loi}. }
\label{tab:jeffreys}
\end{table}

\section{Testing $\mergv$CDM model with Cosmological Data }
\label{sec:results}

In this section, we aim to find observational constraints on the cosmological parameters of the $\mergv$CDM model by considering the priors assumed in Table~\ref {tab:priors}.
We have summarized the observational constraints on parameters of the $\mergv$CDM model in Table~\ref{tab:vcdm-results-1} where we report mean values and 68\% confidence intervals of the parameters from ``CMB'', ``CMB+R21'', ``CMB+SN'', ``CMB+SN+R21'' and ``CMB+BAO+SN (CBS)'' datasets. It is worth mentioning that by using only ``CMB'' data, our model formally reduces the tension in the $H_0$ parameter, but this is due to a volume effect from the increase of the error bars for a larger number of degrees of freedom, and the mean value is actually shifting in the wrong direction.\\

\begin{figure}[ht]
\centering
  \includegraphics[width= .7\linewidth]{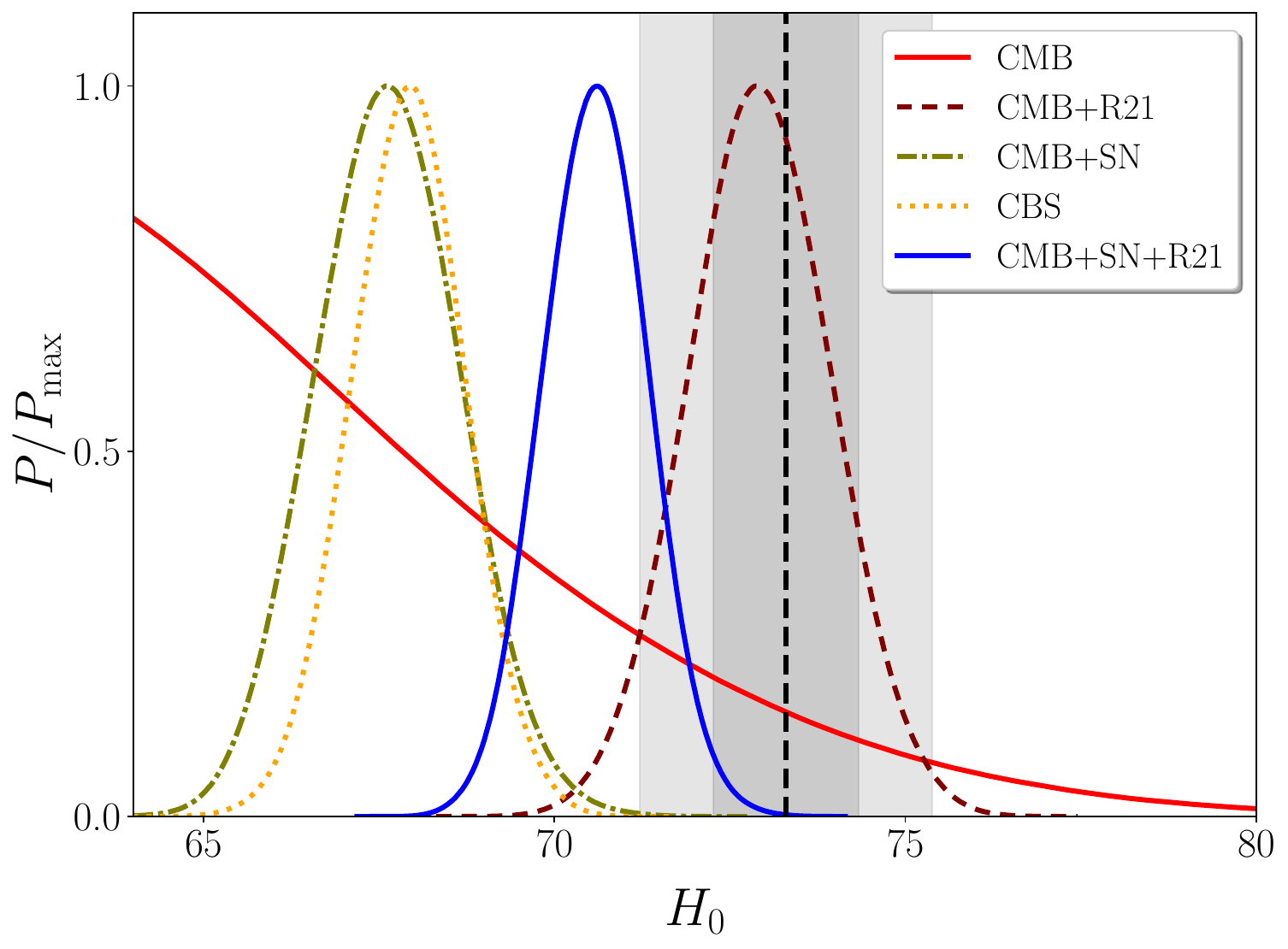}
  \caption{\footnotesize 1D posteriors of $H_0$ are shown for the different dataset combinations. The shaded area shows the measurement of $H_0$ done by the SH0ES team and its $1\sigma$ and $2\sigma$ errors~\citep{Riess:2021jrx}. }
  \label{fig:1D-H0}
\end{figure}

\begin{figure}[ht]
\centering
  \includegraphics[width=.7\linewidth]{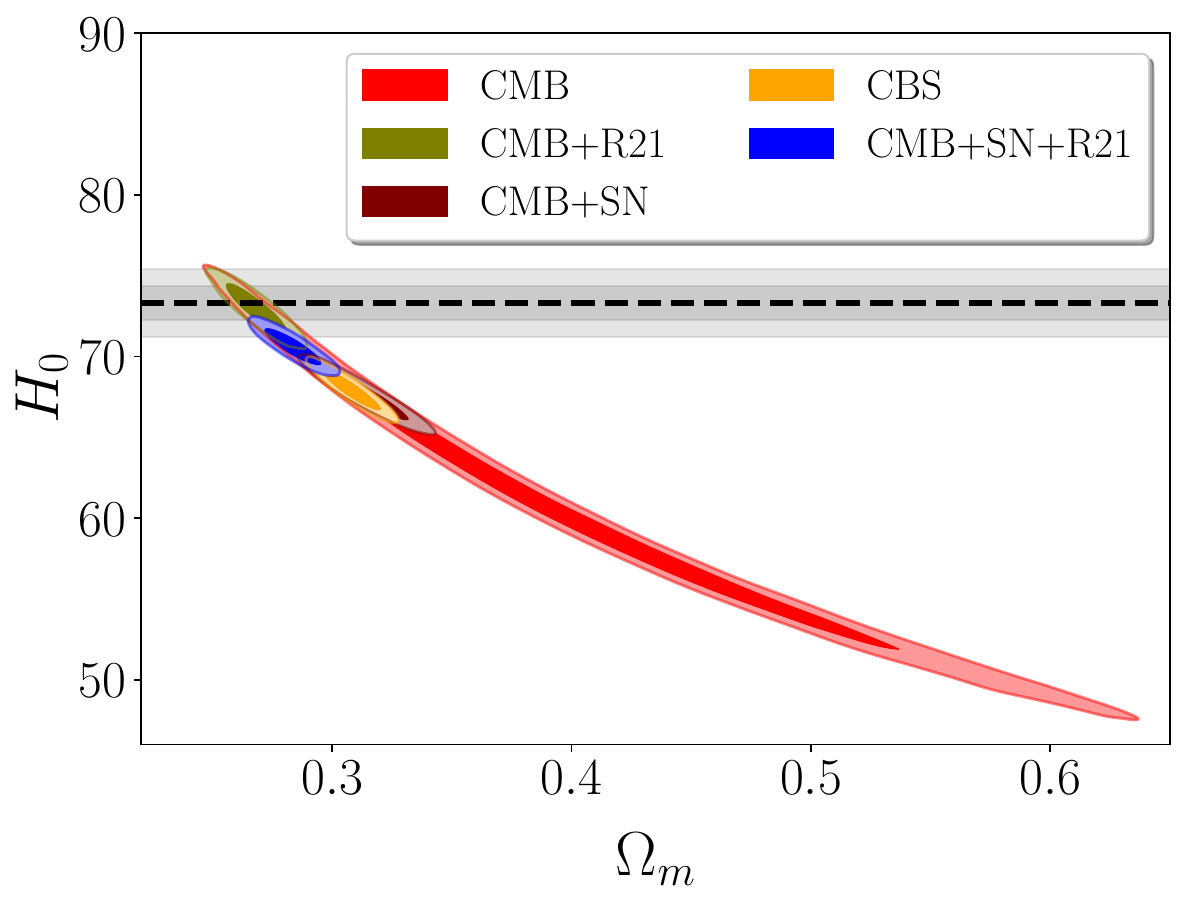}
  \caption{\footnotesize 2D contour plots at $1\sigma$ and $2\sigma$ confidence levels for $H_0$ vs. $\Omega_m$ for the $\mergv$CDM model for different combination of data. The dashed line marker is a central value of the $H_0$ measurement reported by the SH0ES team (R21) and the shaded area shows the $1\sigma$ and $2\sigma$ confidence levels of this measurement. }
  \label{fig:2D-H0-omegam}
\end{figure}

\begin{figure}[ht]
\centering
  \includegraphics[width=0.9\linewidth]{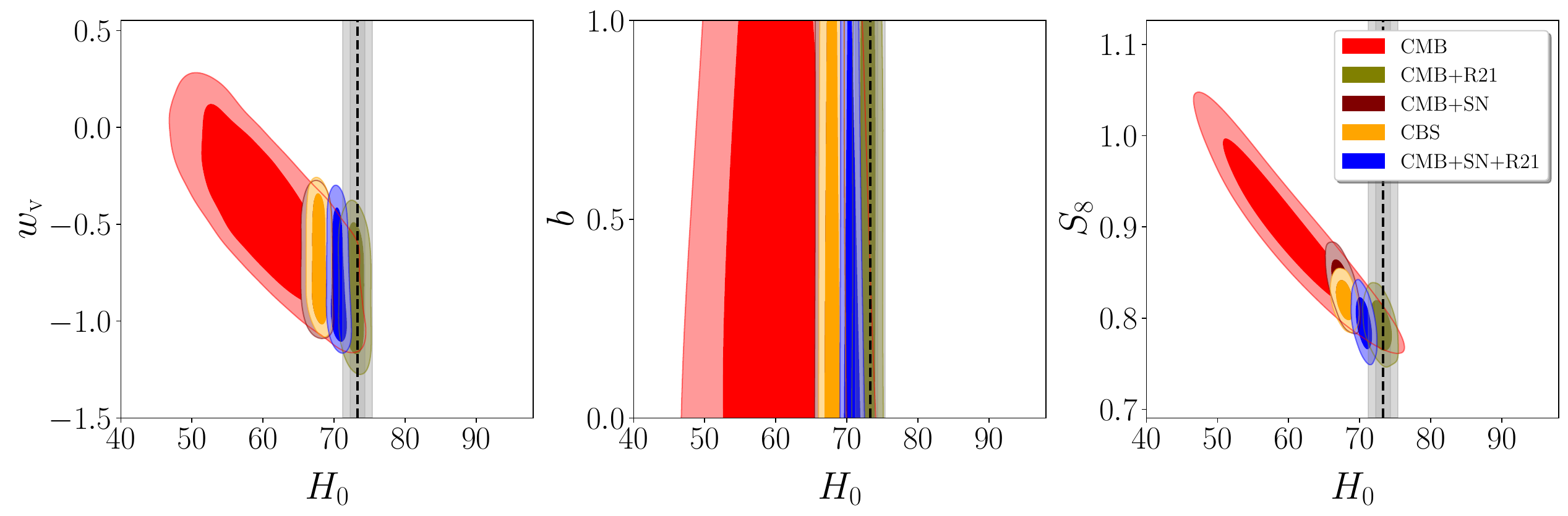}
  \includegraphics[width=0.9\linewidth]{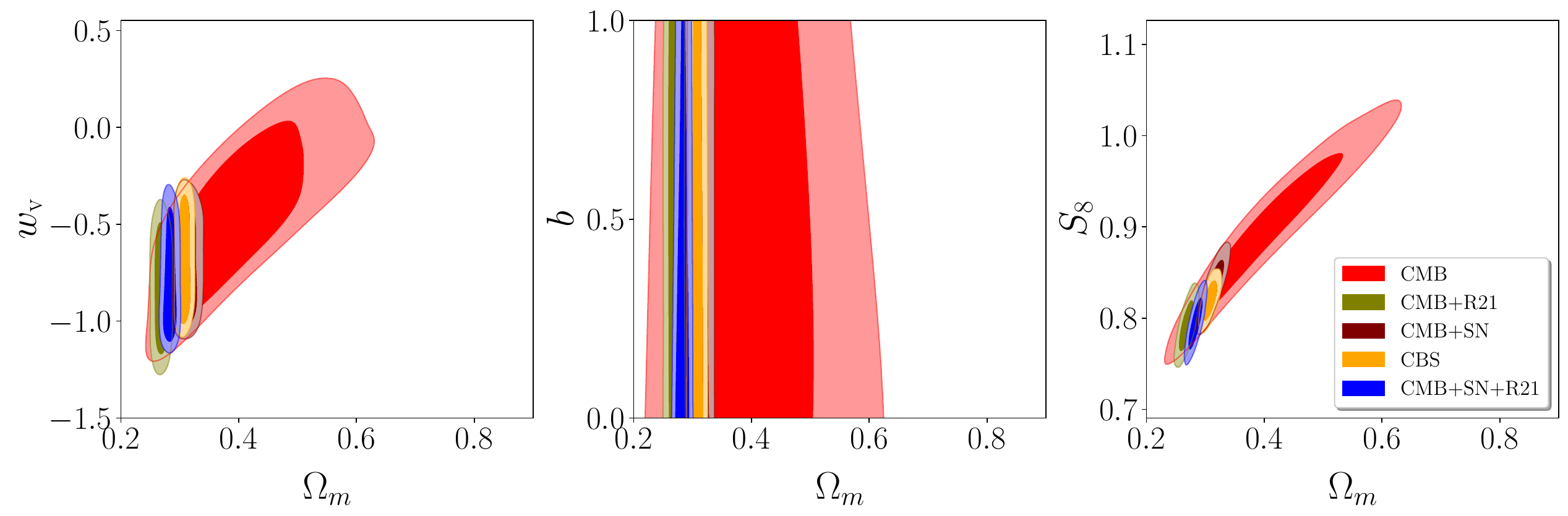}
  \caption{\footnotesize 2D contour plots of $H_0$ and $\Omega_m$  vs. the parameters $w_{\rm v}$ , $b$, and $S_8$ for the $\mergv$CDM model for different combination of data. The vertical dashed line in the upper panel indicates the measurement of the Hubble parameter made by~\citep{Riess:2021jrx} and the shaded area depicts the 1$\sigma$ and 2$\sigma$ error bars.}
  \label{fig:H0-and-omegam-multiple}
\end{figure}

In Fig.~\ref{fig:1D-H0}, we have presented relative 1D posteriors of $H_0$ from different combinations of datasets. The shaded area shows the measurement of $H_0$ done by the SH0ES team~\citep{Riess:2021jrx}. Moreover, in Fig.~\ref{fig:2D-H0-omegam} we see a two-dimensional contour plot of $H_0$ vs. $\Omega_m$ for different combinations of data. As we see considering the R21 prior for the $H_0$ parameter leads to higher values for the Hubble constant but the combination of CMB, BAO, and SN data increases the tension again. Finally, different contour plots at $1\sigma$ and $2\sigma$ confidence levels showing the correlation between $H_0$ and $\Omega_m$ vs. different parameters of the model are shown in Fig.~\ref{fig:H0-and-omegam-multiple}.

\begin{table*}[t]
	\centering
	\begin{tabular} {c||c c c c c c c}
		Parameter & CMB &  CMB+R21 & CMB+SN & CMB+SN+R21 & CBS \\
		\hline
		\hline
		{\boldmath$w_{\rm v}$} & $-0.43\pm 0.30$& $  -0.84^{+0.21}_{-0.25}$ & $-0.71^{+0.18}_{-0.27}$ &$-0.79^{+0.14}_{-0.29}$ & $-0.69^{+0.17}_{-0.28}$ \\
		\hline
		{\boldmath$b $} &$\rm{unconstrained}$  & $\rm{unconstrained}$ & $\rm{unconstrained}$ & $< 0.554$ & $\rm{unconstrained}$ \\
		\hline
		{\boldmath$\Omega_m $} &$0.409^{+0.068}_{-0.10}$ & $0.2684\pm 0.0085$  & $0.314\pm 0.012$ & $0.2836\pm 0.0079$ & $0.3080\pm 0.0079$ \\
		\hline
		{\boldmath$H_0$} & $60.3^{+5.7}_{-7.0}$&$ 72.9\pm 1.0$  & $67.6\pm 1.0$ & $70.59\pm 0.75$ &$ 67.91\pm 0.82$ \\
		\hline
		{\boldmath$S_8 $} & $0.898\pm 0.060$& $0.792\pm 0.019$& $0.833\pm 0.021$  &$0.795\pm 0.019$ & $0.820\pm 0.014$\\
		\hline
		{\boldmath$10^9 A_s $} & $2.092\pm 0.034$& $2.092\pm 0.034$& $2.091\pm 0.033$ & $2.091\pm 0.035$ & $2.091\pm 0.034$ \\
		\hline
		{\boldmath$n_s $} & $0.9618\pm 0.0057$ &$ 0.9649\pm 0.0056$& $0.9630\pm 0.0055$ &$0.9684\pm 0.0055$ & $0.9660\pm 0.0046$ \\
		\hline
		{\boldmath$\tau $} & $0.0518\pm 0.0080$ & $0.0532\pm 0.0081$ &$0.0520\pm 0.0079$ &$0.0544\pm 0.0082$ & $0.0535\pm 0.0080$\\
		\hline
		{\boldmath$\chi^2_{\rm R21}$} & $-- $ & $1.1$ & $--$ &$7.3$ & $--$ \\
		\hline

		{\boldmath$\chi^2_{\rm BAO}$} & $-- $ & $--$ & $--$ & $--$ &$6.3$ \\
		\hline
		{\boldmath$\chi^2_{\rm SN}$} & $-- $ & $--$ & $1035.8$ & $1040.7$ &$1035.6$ \\
		\hline
		{\boldmath$\chi^2_{\rm CMB}$} & $630.2 $ & $633.4$ & $630.8$ & $632.8$ & $630.7$ \\
		\hline
		\hline
		{\boldmath$\chi^2_{\rm ToT}$} &$630.2 $ & $634.5$ & $1666.6$ & $1680.8$ & $1672.6$ \\
		\hline
		\hline
		{\boldmath$\Delta \rm AIC $} & $+3.9$ &$ -13.1$ & $+4.69$ & $-2.06$ &$ +5.07$  \\
		\hline
		\hline	
          {\boldmath$\rm \ln \mathcal{B}_{12}$} & $-0.26$ & $+6.08$ & $-2.39$ &$+0.20$ & $-2.86$\\ 
          \hline
          \hline
	\end{tabular}
	\caption{\label{tab:vcdm-results-1}  \footnotesize{68 $\%$ CL constraints on free and derived parameters of the $\mergv$CDM model from \emph{Planck} CMB power spectra in combination with BAO and Type Ia SuperNovae data plus a Gaussian prior on the Hubble constant $H_0$ as measured by SH0ES team.}}
\end{table*}

In the last row of Table~\ref{tab:vcdm-results-1} we have shown the values of Bayes Factor in the logarithmic scale $\ln \mathcal{B}_{12}$ computed for $\mergv$CDM model against the $\Lambda$CDM one considering all the datasets. We find that the values of $\ln \mathcal{B}_{12}$ for ``CMB+SN'' and a combination of ``CMB+BAO+SN'' are negative indicating that $\Lambda$CDM is preferred in these datasets significantly and strongly, respectively. However, in the combination of CMB data with a prior on $H_0$ (R21) we find a positive value for $\ln \mathcal{B}_{12}$ that indicates very strong evidence for $\mergv$CDM model over the reference model $\Lambda$CDM.
This is because for the $\mergv$CDM model the Hubble tension is reduced because of a volume effect, so it is easier to accommodate a higher $H_0$ value than in the standard $\Lambda$CDM scenario. Finally, for the ``CMB'' and ``CMB+SN+R21'' cases, we find a value of $|\ln \mathcal{B}_{12}|$ that is $<1$ indicating that the $\mergv$CDM model is as favored as $\Lambda$CDM from the data point of view.\\

\begin{figure}[ht]
\centering
  \includegraphics[width=0.7\linewidth]{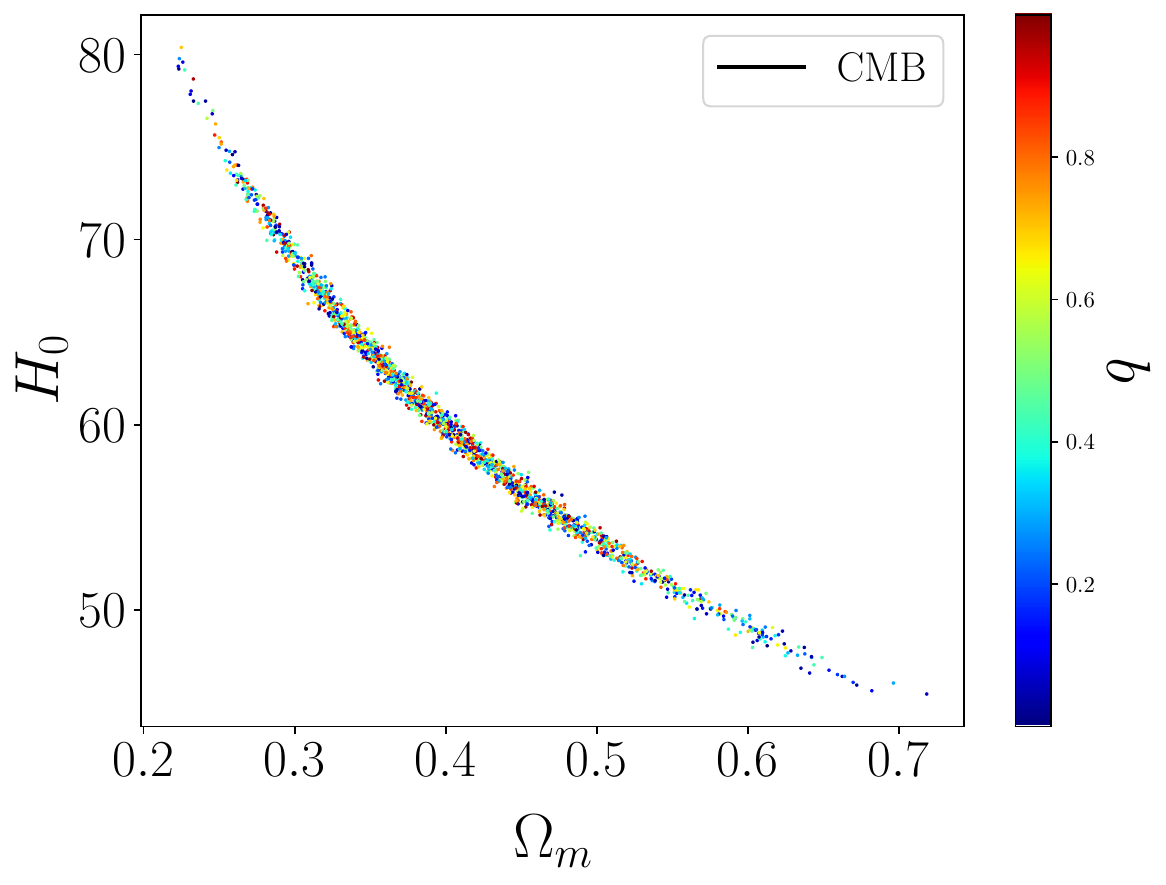}
    \includegraphics[width=0.7\linewidth]{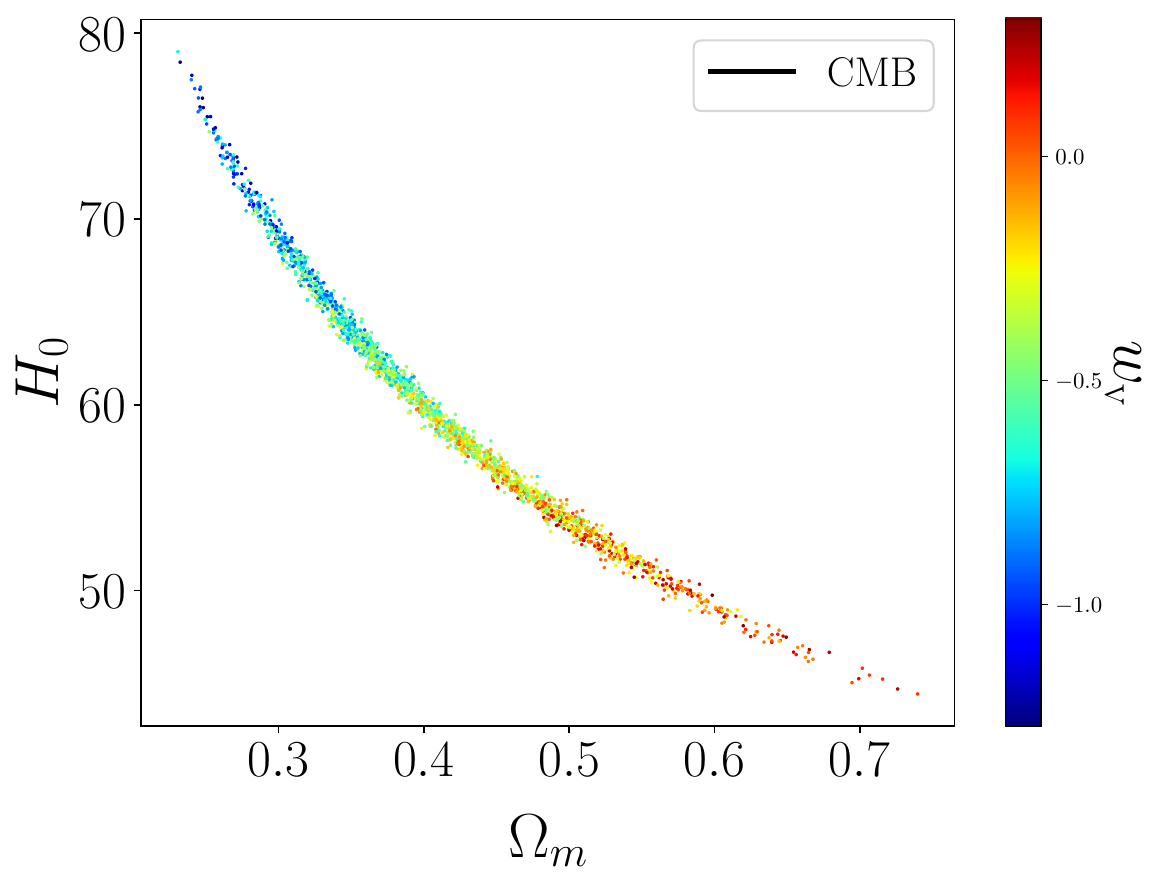}
  \caption{\footnotesize \textit{Upper:} Three-dimensional contour of $H_0$ vs. $\Omega_m$, where the $b$ parameter is shown as a third dimension by variation of its color. \textit{Lower:} Similar plot as in the upper panel where the effect of $w_{\rm v}$ parameter is shown as the third dimension.}
  \label{fig:3D-omegam-H0-b-w1}
\end{figure}

For more insight we have plotted three-dimensional contours (see Fig.~\ref{fig:3D-omegam-H0-b-w1}) for Hubble constant $H_0$ vs. $\Omega_m$, in one case with the merging rate of cosmic voids $b$, as the third dimensions in color, and in the other case with the equation of state parameter $w_{\rm v}$. Fig.~\ref{fig:3D-omegam-H0-b-w1} shows that while there is no clear correspondence between larger values of the Hubble expansion rate and $b$, on the contrary, they occur with $w_{\rm v}$ closer to $-1$.

\begin{figure*}[ht]
	\centering
	\includegraphics[width=\linewidth]{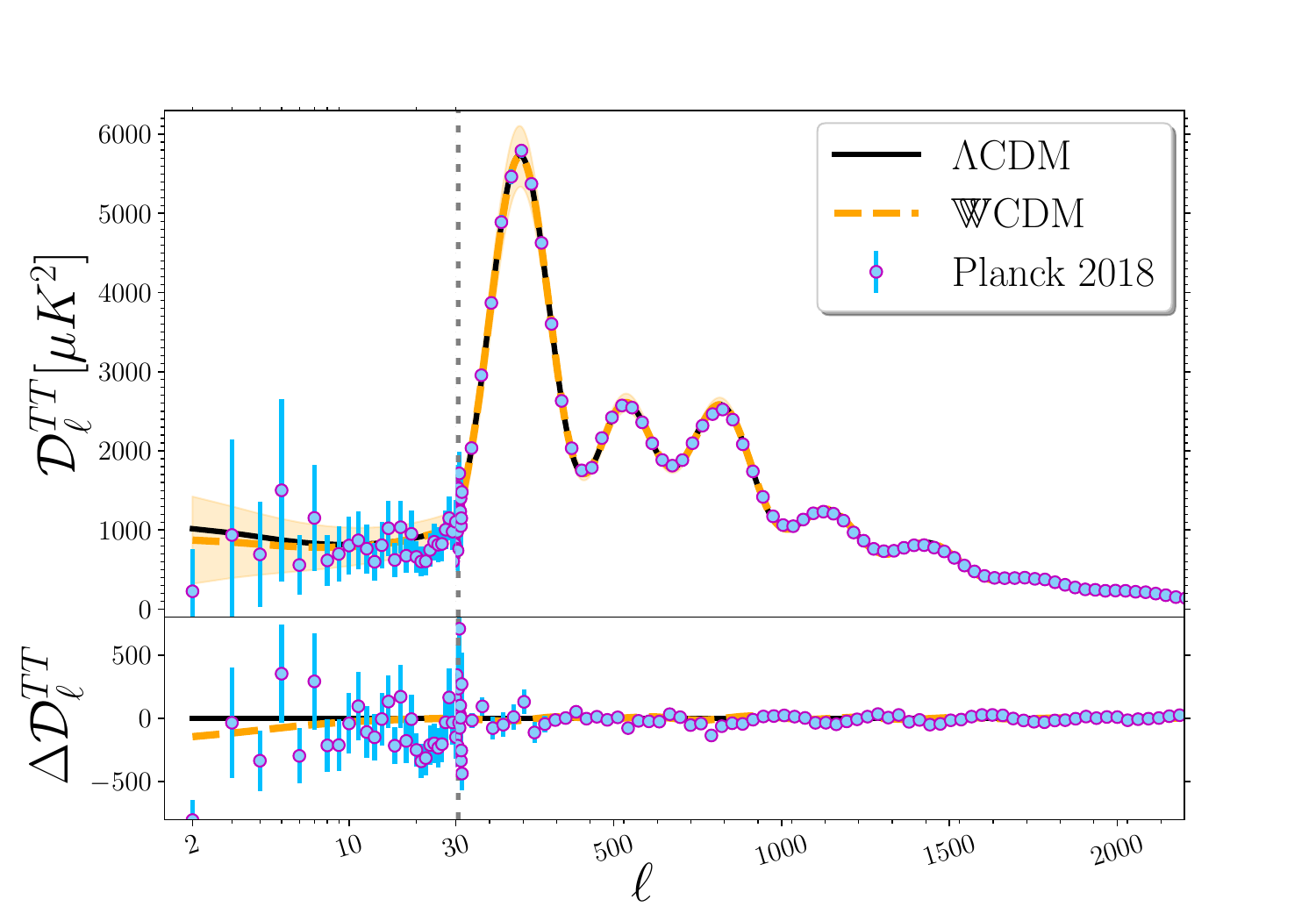}
	\caption{\footnotesize Power spectrum of the temperature anisotropies for the $\mergv$CDM model with a dashed orange line obtained with the best-fit of the parameters for the ``CMB''. The theoretical prediction of the $\Lambda \rm{CDM}$ model is shown with a solid black line. Blue points show observational data of \emph{Planck} 2018 release. The shaded area represents the cosmic variance of the power spectrum. The lower panel shows the residual of the two models to observational data.}
	\label{fig:CMB-powerspectrum}
	\end{figure*}

We calculated the power spectrum of the CMB temperature anisotropies for the $\mergv$CDM model for the best fit obtained with the ``CMB'' dataset. The result is shown in Fig.~\ref{fig:CMB-powerspectrum} which demonstrates a very good consistency with observational data. The shaded area in Fig.~\ref{fig:CMB-powerspectrum} denotes the cosmic variance of the power spectrum. According to this figure for higher multipoles ($\ell$) the prediction of $\mergv$CDM and $\Lambda \rm{CDM}$ are very similar. However, for the plateau part of the power spectrum ($\ell < 30$) the tail of the $\mergv$CDM model is lower than the standard model, helping in recovering the low quadruple. It shows our model has a weaker Integrated Sachs-Wolfe effect (ISW) which can be due to the effect of large over-densities in the cosmic super-clusters. This feature is expected because of the large value of $S_8$ parameter (according to Table~\ref{tab:vcdm-results-1}) which means we have more clusters in $\mergv$CDM model than $\Lambda$CDM. \\

\begin{figure}[ht]
	\centering
	\includegraphics[width=.7\linewidth]{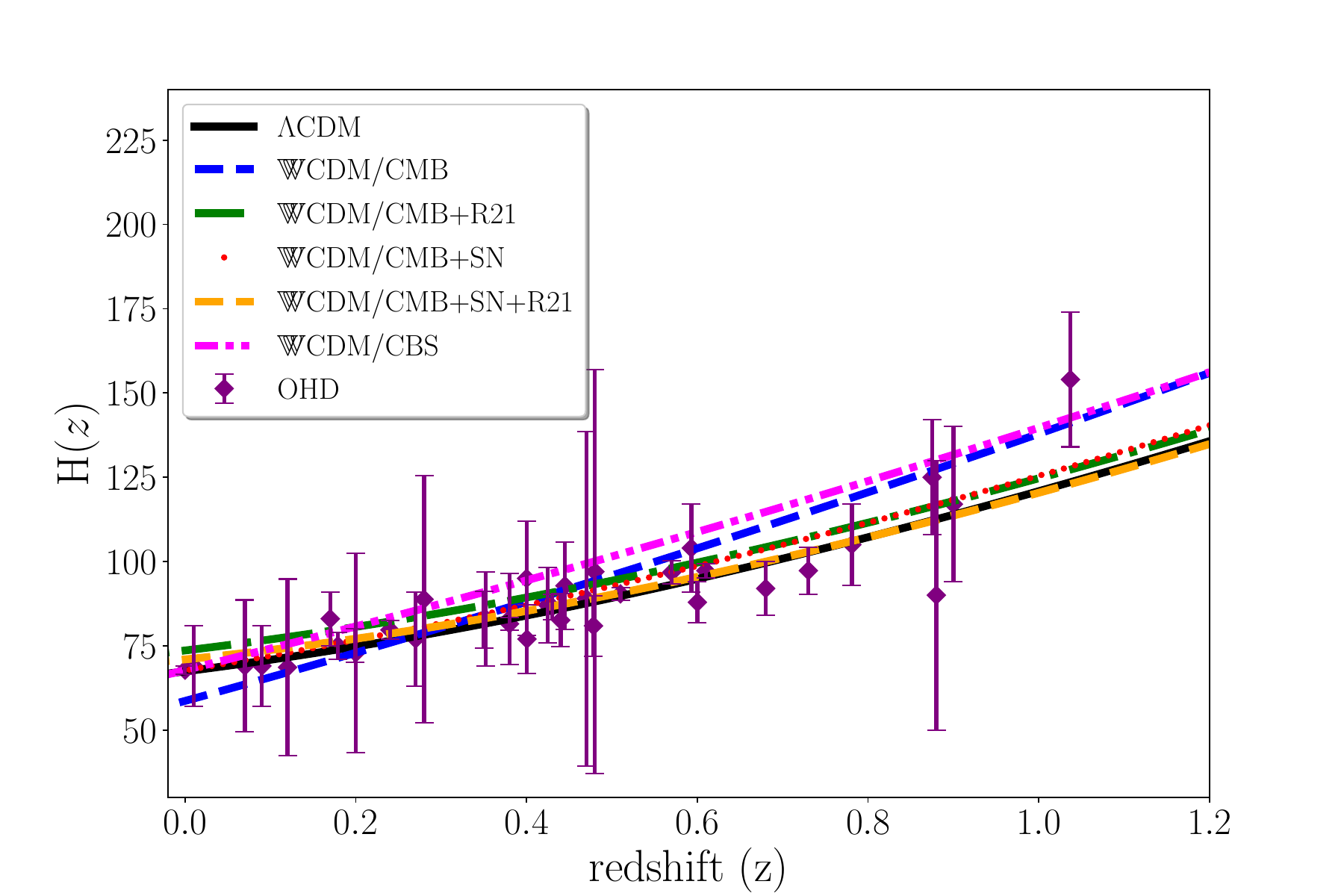}
	\caption{\footnotesize Evolution of Hubble parameter for the $\mergv$CDM model, using best-fit values of cosmological parameters for ``CMB'' (dashed blue), ``CMB+R21'' (dash-dotted green), ``CMB+SN'' (dotted red), ``CMB+SN+R21'' (dashed orange), and ``CBS'' (dotted magenta). The $H(z)$ of $\Lambda \rm{CDM}$ model is shown with a solid black line. OHD points in Table~\ref{tab:OHD-data} are shown in purple.}
	\label{fig:hz_data}
\end{figure}

\begin{figure}[ht]
	\centering
	\includegraphics[width=.7\columnwidth]{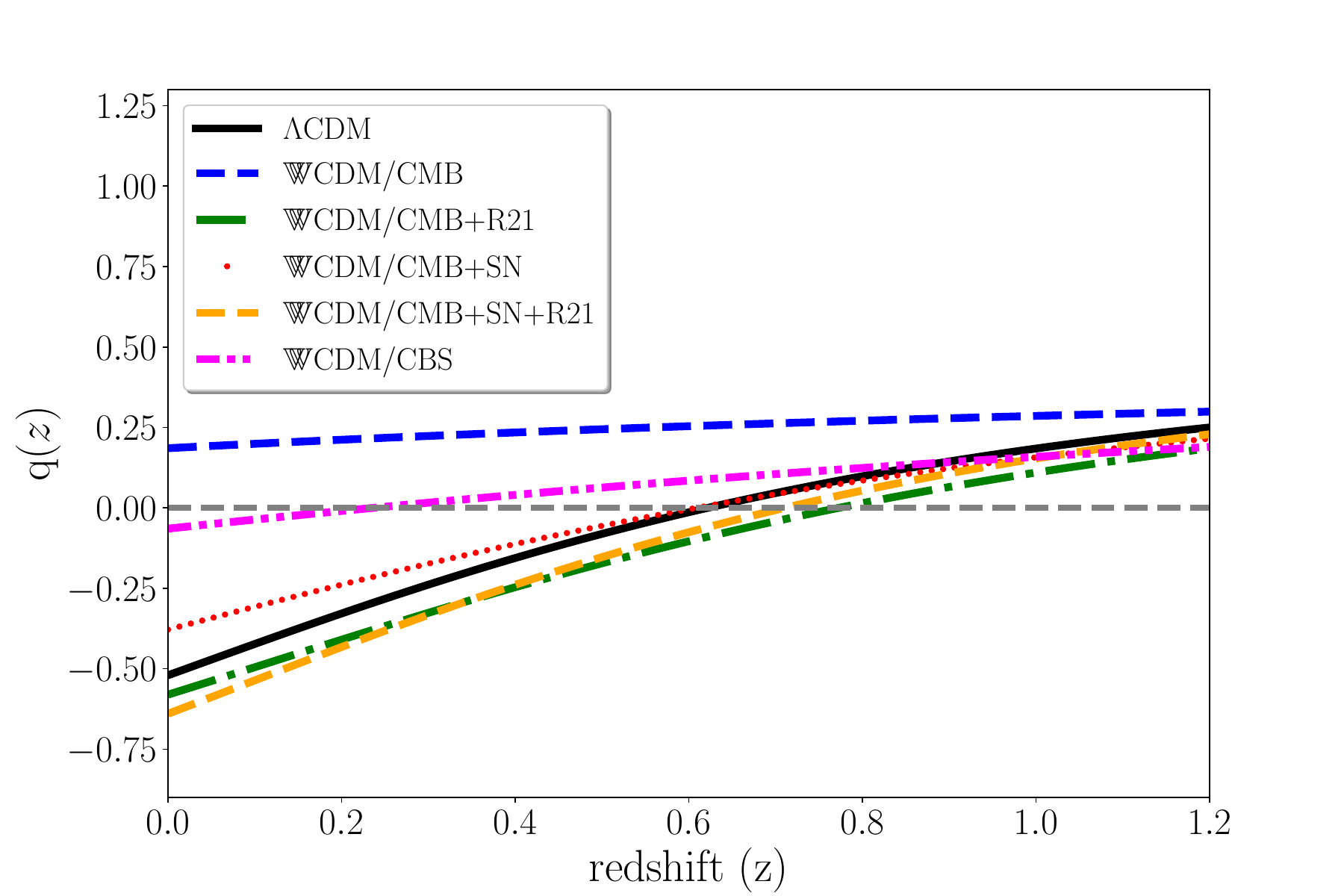}
	\caption{\footnotesize Evolution of deceleration parameter $q(z)$ for the $\mergv$CDM model, using best-fit cosmological parameters from Table~\ref{tab:vcdm-results-1} for ``CMB'' (dashed blue), ``CMB+R21'' (dash-dotted green), ``CMB+SN'' (dotted red), ``CMB+SN+R21'' (dashed orange), and ``CBS'' (dotted magenta). For comparison, the standard $\Lambda \rm{CDM}$ prediction is also shown in a solid black line. }  
	\label{fig:qz}
\end{figure}

In Fig.~\ref{fig:hz_data} we see the evolution of the Hubble parameter for the $\mergv$CDM model based on the best-fit of the different dataset combinations in comparison with $\Lambda \rm{CDM}$ model. The data points include the 
current Observational Hubble Data (OHD) in the ranges $0 \leq z \leq 2.36$ (see Table~\ref{tab:OHD-data}), which were obtained by using the ``cosmic chronometric approach (CCA)''~\citep{Farooq:2016zwm, Farooq:2013syn, Moresco:2012jh, Simon:2004tf}. As we can see in Fig.~\ref{fig:hz_data} the $\mergv$CDM model at low redshift shows good consistency with observational data. However, at higher redshifts, the Hubble rate based on best-fit parameters from the analysis of only ``CMB'' and also ``CMB+BAO+SN (CBS)`` data show deviation from some observational data points.
Moreover, for investigating the accelerating behavior of $\mergv$CDM model, we have represented the evolution of the deceleration parameter $q(z)$ in Fig.~\ref{fig:qz}. \\

\begin{figure}[ht]
	\centering
	\includegraphics[width=.7\linewidth]{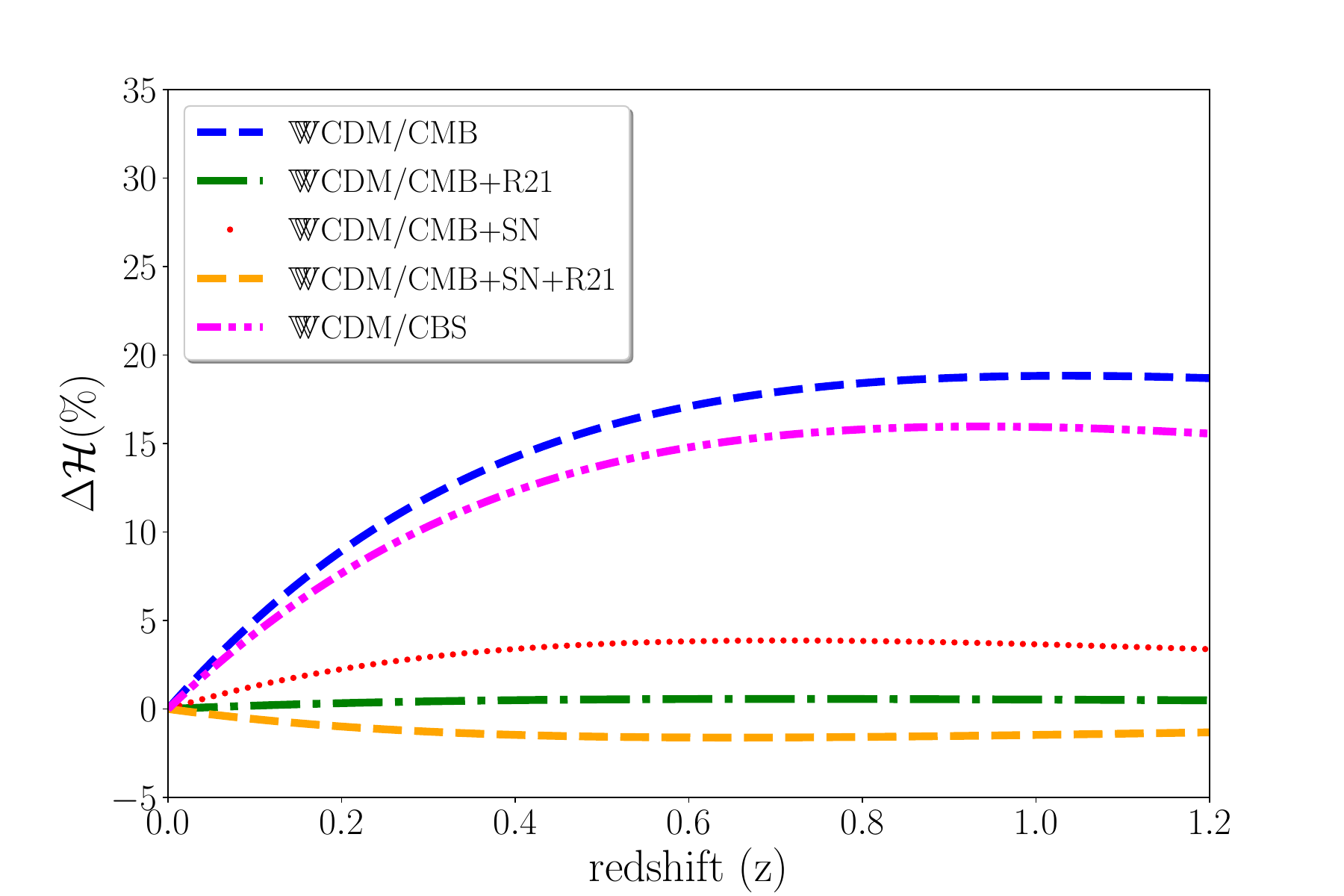}
	\caption{\footnotesize Relative difference of the Hubble parameter as a function of the redshift for the $\mergv$CDM model, using best-fit cosmological parameters from Table~\ref{tab:vcdm-results-1} for ``CMB'' (dashed blue), ``CMB+R21'' (dash-dotted green), ``CMB+SN'' (dotted red), ``CMB+SN+R21'' (dashed orange), and ``CBS'' (dotted magenta). }
	\label{fig:deltaH}
\end{figure}

The relative difference in Hubble parameter of the $\mergv$CDM model with $\Lambda$CDM model is defined as follows:
\begin{equation}
\Delta \mathcal{H}(z)=100 \times \left[ \frac{\mathcal{H}_{\mergv \rm{CDM}}(z)}{\mathcal{H}_{\Lambda \rm{CDM}}(z)} -1 \right].
\end{equation}
The above quantity is illustrated in Fig.~\ref{fig:deltaH} as a function of redshift.

\begin{table*}[]
	\centering
	\begin{tabular}{c c c c c c}
		\hline
		\hline
		$z$  & $H(z)$ & Ref. & $z$  & $H(z)$ & Ref.  \\
		\hline
		
		$0.0$&  $67.77 \pm 1.3$ &\citep{DES:2018rjw} & $0.07$&$69.0 \pm 19.6$ &\citep{Stern:2009ep} \\
		$0.09$ & $69.0 \pm 12.0$ &~\citep{Simon:2004tf} & $0.01$ & $69.0 \pm 12.0$ &~\citep{Stern:2009ep} \\
		$0.12$ & $68.6 \pm 26.2$ &~\citep{Stern:2009ep} & $0.17$ & $83.0 \pm 8.0$ &~\citep{Stern:2009ep} \\
		$0.179$ & $75.0 \pm 4.0$ &~\citep{Moresco:2012jh} & $0.1993$ & $75.0 \pm 5.0$ &~\citep{Moresco:2012jh} \\
		$0.2$ & $72.9 \pm 29.6$ &~\citep{Stern:2009ep} & $0.24$ & $79.7 \pm 2.7$ &~\citep{Gaztanaga:2008xz} \\
		$0.27$ & $77.0 \pm 14.0$ & ~\citep{Stern:2009ep} & $0.28$ & $88.8 \pm 36.6$  &\citep{Stern:2009ep} \\
		$0.35$ & $82.7 \pm 8.4$ &~\citep{BOSS:2016wmc} & $0.352$ & $83.0 \pm 14.0$ &~\citep{Moresco:2012jh} \\
		$0.38$ & $81.5 \pm 1.9$ &~\citep{Chuang2013} & $0.3802$ & $83.0 \pm 13.5$ &~\citep{BOSS:2016wmc} \\
		$0.4$ & $ 95.0 \pm 17$ & ~\citep{Simon:2004tf} & $0.4004$ & $77.0 \pm 10.2$ &~\citep{Moresco:2016mzx} \\
		$0.4247$ & $87.1 \pm 11.2$ &~\citep{Moresco:2016mzx} & $0.43$ & $86.5 \pm 3.7$ &~\citep{Gaztanaga:2008xz} \\
		$0.44$ & $82.6 \pm 7.8$ &~\citep{Blake2012} & $0.44497$ & $92.8 \pm 12.9$ & ~\citep{Moresco:2016mzx} \\
		$0.47$ & $ 89 \pm 49.6$ &~\citep{Ratsimbazafy:2017vga} & $0.4783$ & $80.9 \pm 9.0$ &~\citep{Moresco:2016mzx} \\
		$0.48$ & $97.0 \pm 60.0$ &~\citep{Stern:2009ep} & $0.51$ & $90.4 \pm 1.9$ &~\citep{Chuang2013} \\
		$0.57$ & $96.8 \pm 3.4$ &~\citep{Sarmah:2022hmf} & $0.593$ & $104 \pm 13$ &~\citep{Moresco:2012jh} \\
		$0.60$ & $87.9 \pm 6.1$ & ~\citep{Blake2012} & $0.61$ & $97.3 \pm 2.1$ &~\citep{Chuang2013} \\
		$0.68$ & $92.0 \pm 8.0$ & ~\citep{Moresco:2012jh} & $0.73$ & $97.3 \pm 7.0$ & ~\citep{Blake2012} \\
		$0.781$ & $105 \pm 12.0$ &~\citep{Moresco:2012jh} & $0.875$ & $125 \pm 17.0$ &~\citep{Moresco:2012jh} \\
		$0.88$ & $90 \pm 40.0$ & ~\citep{Stern:2009ep} & $0.9$ & $117 \pm 23.0$ &~\citep{Stern:2009ep} \\
		$1.037$ & $154 \pm 20.0$ &~\citep{Gaztanaga:2008xz} & $1.3$ & $168 \pm 17$ &~\citep{Stern:2009ep} \\
		$1.363$ & $160 \pm 33.6$ &~\citep{Moresco:2015cya} & $1.43$ & $177 \pm 18$ &~\citep{Stern:2009ep} \\
		$1.53$ & $140 \pm 14$ &~\citep{Stern:2009ep} & $1.75$ & $202 \pm 40$ &~\citep{Moresco:2015cya} \\
		$1.965$ & $186.5 \pm 50.4$ &~\citep{Gaztanaga:2008xz} & $2.3$ & $224 \pm 8.0$ &~\citep{Busca2013} \\
		$2.34$ & $222 \pm 7.0$ &~\citep{BOSS:2014hwf} & $2.36$ & $226 \pm 8$ &~\citep{BOSS:2013igd} \\
		\hline
		
	\end{tabular}

	\caption{\footnotesize The currently available OHD measurements.}

	\label{tab:OHD-data}
\end{table*}


\section{Summary and Discussions}
\label{sec:Discussion}

In this paper, we have investigated the observational predictions of a homogeneous and isotropic cosmological fluid with a non-linear EoS given by Eq.~\eqref{P_v} for DE. The model introduces two additional free parameters beyond the $\Lambda$CDM model. In the standard cosmological model, the dark fluid is attributed to the accelerating expansion of the Universe, characterized by $w_{\rm v} \simeq -1$ and $\Omega_{\rm v} \simeq 0.7$.

We examined an effective fluid with a non-linear quadratic EoS that drives the accelerating expansion. Such non-linear terms in the EoS may arise from the merger process of clusters/voids. The outcome of such mergers could be the observed accelerating expansion at cosmic scales.

In this study, we considered Einstein's field equation at the background level in the presence of the dark fluid to investigate the Hubble parameter tension. Additionally, we calculated the amplitude of matter density fluctuations at the perturbation level.
We assessed the model's consistency with observational data by examining it with various combinations of cosmological data. Testing the model solely with ``CMB'' observational data (Planck 2018), we obtained a constraint for the Hubble constant that reads $H_0=60.3^{+5.7}_{-7.0} \ \rm{km/s Mpc}$. This result suggests a reduction in the Hubble tension to approximately a $2 \sigma$ level. However, due to significant large error bars in the $H_0$ parameter caused by a volume effect, the Hubble parameter lacks precise constraints.\\

Furthermore, we demonstrated the consistency of the Hubble rate as a function of redshift in our model compared to OHD from Table~\ref{tab:OHD-data}, contrasting it with the Standard $\Lambda \rm{CDM}$ model in Fig.~\ref{fig:hz_data}. Additionally, we explored the dynamical behavior of the model by calculating other observational functions, such as the deceleration parameter (Fig.~\ref{fig:qz}), and relative Hubble parameter~(Fig.~\ref{fig:deltaH}).\\
We calculated the Bayesian Evidence for our model in comparison with the reference model $\Lambda$CDM (last row of Table~\ref{tab:vcdm-results-1}). Our findings indicate a strong preference for the $\mergv$CDM model over the $\Lambda$CDM model when utilizing CMB (\emph{Planck}) data along with a prior on $H_0$ (R21). However, this preference is attributed to a reduced Hubble constant tension in this specific scenario and diminishes when R21 is not taken into account. It should be mention that the combination of CMB, BAO and SNeIa data doesn't lead to improvement in Hubble tension problem, similar to $\Lambda$CDM. This model can't resolve the Hubble tension completely.

It is important to note that our study focused on a specific case of a non-linear EoS. While our analysis reveals favorable behavior in various observational tests, it is noteworthy that the well-known cosmological tensions persist when the full dataset combination is explored. Future work will be directed towards considering more general non-linear EoS for dark energy. In subsequent studies, we plan to explore these cases for the dark energy model and also investigate the non-linear structure formation of such models to derive statistics for dark matter halos and cosmic voids.

\section*{Acknowledgments}
The authors are grateful to Hassan Firouzjahi for his valuable remarks and comments. Also, we thank Sunny Vagnozzi for his helpful comments. We acknowledge Haidar Sheikhahmadi for his revision and fruitful comments. A. T. would like to thank the Yukawa Institute for Theoretical Physics at Kyoto University, ICTP, University of Rwanda, and ICTP-EAIFR for their kind hospitalities when some parts of the project were carried out.
E.D.V. is supported by a Royal Society Dorothy Hodgkin Research Fellowship. E.Y. is supported by the Islamic Azad University, Ayatollah Amoli branch, Amol, Iran. This article is based upon work from COST Action CA21136, Addressing observational tensions in cosmology with systematics and fundamental physics (CosmoVerse) supported by COST (European Cooperation in Science and Technology).


\vspace{0.7cm}

\appendix

\section{Quadratic Equation of State for Evolving Merging Fluid}
In our model, we have considered the merging of similar objects, leading to the formation of larger structures. 
For example, two identical clusters with number density $n_{\rm clu}$ can merge to form a super-cluster with number density $n_{\rm scl}$. Moreover, voids can also merge to create a super-void.
Expanding super-voids can reduce the surface tension of super-clusters located on their borders, potentially facilitating the division of super-clusters into individual clusters. Considering the simultaneous merger and expansion of voids alongside clusters in the cosmic web, the merger process of clusters can be described as a chemical reaction in thermal equilibrium as:
\begin{equation} 
	n_{\rm clu} + n_{\rm clu} \rightleftharpoons n_{\rm scl} \,.
\end{equation}
In the following we define an equilibrium constant $K$, similar to the same concept in chemical interaction for fluids, to be
\begin{equation}
	\label{cvt1}
	K=\frac{n_{\rm scl}}{n^2_{\rm clu}}\,.
\end{equation}
For an ideal gas law $P = nk_{B}T$, the pressure after the merger process ($P_{\rm aft}$) can be related to the pressure before merging ($P_{\rm bef}$) as \cite{Mohammadi:2023idz}
\begin{equation}
	\label{cvt2}
	P_{\rm aft}=P_{\rm bef}-\frac{P^2_{\rm bef}}{(k_{B}T)}K \,.
\end{equation}
It means the merger process leads to a reduction in primary pressure ($P_{\rm bef}$), acting akin to a \textit{negative pressure}. Considering the pressure for our model containing voids before merger as $p_{\rm bef}=w\rho_{\rm v}$, then after of merging process, the pressure is given by \eqref{P_v}.

\end{document}